\documentclass[smallcondensed]{svjour3}
\usepackage{amssymb}
\usepackage{float}
\usepackage{graphicx}
\usepackage{amsmath}
\usepackage{multirow}
\usepackage{subfig}
\usepackage{algpseudocode}
\usepackage{array}
\newcolumntype{P}[1]{>{\centering\arraybackslash}p{#1}}\usepackage[linesnumbered,ruled]{algorithm2e}

\newcommand{\bm}{\mathbf}
\newcommand{\alg}{{\sc DyLink2Vec}}

\begin{document}
\title{\alg: Effective Feature Representation for \\Link Prediction in Dynamic Networks}
\titlerunning{\alg}       

\author{Mahmudur Rahman         \and
        Tanay Kumar Saha  \and 
        Mohammad Al Hasan \and 
        Kevin S. Xu \and 
        Chandan K. Reddy}


\institute{Mahmudur Rahman, Tanay Kumar Saha, Mohammad Al Hasan \at
              Indiana University Purdue University Indianapolis \\
              \email{\{mmrahman, tksaha, alhasan\}@iupui.edu}           
           \and
           Kevin S. Xu \at
              University of Toledo \\
              \email{kevin.xu@utoledo.edu}
           \and 
           Chandan K. Reddy \\
           Virginia Tech \\
           \email{reddy@cs.vt.edu}
}

\date{Received: date / Accepted: date}

\maketitle

\begin{abstract}
The temporal dynamics of a complex system such as a social network or a
communication network can be studied by understanding the patterns of link
appearance and disappearance over time. A critical task along this
understanding is to predict the link state of the network at a future time
given a collection of link states at earlier time points. In existing literature,
this task is known as link prediction in dynamic networks. Solving this task is
more difficult than its counterpart in static networks because an effective
feature representation of node-pair instances for the case of
dynamic network is hard to obtain. 
To overcome this problem, we propose a
novel method for metric embedding of node-pair
instances of a dynamic network. The proposed method models the metric embedding
task as an optimal coding problem where the objective is to minimize the
reconstruction error, and it solves this optimization task using a gradient
descent method. We validate the effectiveness of the learned feature
representation by utilizing it for link prediction in various real-life
dynamic networks. Specifically, we show that our proposed link prediction
model, which uses the extracted feature representation for the training
instances, outperforms several existing methods that use well-known
link prediction features.
\end{abstract}

\section{Introduction}\label{sec:intro}
Understanding the dynamics of an evolving network is an important research
problem with numerous applications in various fields, including social network
analysis~\cite{Otte:2002}, information retrieval~\cite{jansen:2010},
recommendation systems~\cite{Ricci:2010}, 
and bioinformatics~\cite{jacobs:2001}. A key task
towards this understanding is to predict the likelihood of a future association
between a pair of nodes, having the knowledge about the current state of the network. This task is
commonly known as the {\em link prediction} problem. Since, its formal
introduction to the data mining community by Liben-Nowell et
al.~\cite{Liben.Nowell:2007} about a decade ago, this problem has been studied
extensively by many researchers from a diverse set of disciplines
\cite{Miller.Jordan.ea:09,Ben:2003,Lichtenwalter:2010,Barbieri:2014}.
Good surveys \cite{Hasan:2011,Wang:2015} on link prediction methods are
available for interested readers. 

The majority of the existing works on link prediction consider a static
snapshot of the given network, which is the state of the networks at a given
time \cite{Miller.Jordan.ea:09,Lichtenwalter:2010,Liben.Nowell:2007,Menon:2011}. Nevertheless, for many networks, additional
temporal information such as the time of link creation and deletion is
available over a time interval; for example, in an on-line social or a
professional network, we usually know the time when two persons have become friends;
for collaboration events, such as, a group performance or a collaborative
academic work, we can extract the time of the event from an event calendar.
The networks built from such data can be represented by a {\em dynamic
network}, which is a collection of temporal snapshots of the network. The
link prediction\footnote{Strictly speaking, 
this task should be called as \textit{link forecasting}
since the learning model is not trained on the links at time $t$;
however, we refer it as link prediction due to the popular usage of this term
in the data mining literature.} task on such a network is defined as follows: {\em for a given
pair of nodes, predict the link probability between the pair at time $t+1$ by
training the model on the link information at times $1, 2, \cdots,
t$}. 

\begin{figure}[t!]
  \begin{center}
    \includegraphics[width=0.85\textwidth]{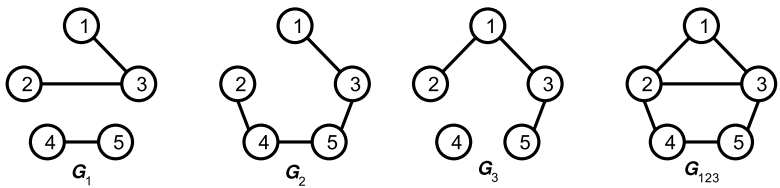}
  \end{center}
  \caption{A Toy Dynamic Network. $G_1$, $G_2$ and $G_3$ are three snapshots of the Network. $G_{123}$ is constructed by superimposing $G_1$, $G_2$ and $G_3$.}
  \label{fig:StaticLimitaion}
\end{figure}

Link prediction methods for static networks fail to take advantage of the
temporal link formation patterns that are manifested by the sequence of
multiple temporal snapshots.  For illustration, let us consider a toy dynamic
network having three temporal snapshots $G_1$, $G_2$ and $G_3$ (see Figure
\ref{fig:StaticLimitaion}). A static link prediction which only considers the
latest time stamp $G_3$ forfeits the temporal signals that are available from
prior snapshots $G_1$ and $G_2$.  Thus, it is oblivious of the fact that the
edge $(4,5)$ once existed. On the other hand, if the static link prediction
method runs on a superposition \cite{Lichtenwalter:2010} of all the available
snapshots ($G_{123}$), it fails to preserve the temporal variation in the
dataset. For example, the superimposed static snapshot fails to distinguish the
link strength between the edges $(3,5)$ and $(4,5)$---even though both edges
appear twice in $G_1$, $G_2$ and $G_3$, the recency of $(3,5)$ may make it more
likely to re-appear than $(4,5)$.

A key challenge of link prediction in a dynamic setting is to find a suitable
feature representation of the node-pair instances which are used for training
the prediction model. For the static setting, various topological metrics (common
neighbors, Adamic-Adar, Jaccard's coefficient) are used as features, but they cannot
be extended easily for the dynamic setting having multiple snapshots of the
network.  In fact, when multiple (say $t$) temporal snapshots of a network are
provided, each of these scalar features becomes a $t$-size sequence. 
Flattening the sequence into a $t$-size vector
distorts the inherent temporal order of the features. G{\"u}ne{\c{s}} et al.~\cite{Gunes2015}
overcome this issue by modeling a collection of time series, each for one of
the topological features. But such a model fails to capture
signals from the neighborhood topology of the edges. There exist few other
works on dynamic link prediction, which use probabilistic
(non-parametric) and matrix factorization based models. These works consider a
feature representation of the nodes and assume that having a link from one
node to another is determined by the combined effect of all pairwise node feature
interactions~\cite{ICML2012Sarkar_828,KevinS:2013,Dunlavy:2011}.
While this is a reasonable assumption to make, the accuracy of such models are highly
dependent on the quality and availability of the node features, as well as the validity of the
above assumption. 

There exist a growing list of recent works which use unsupervised methodologies
for finding metric embedding of nodes in a graph~\cite{Perozzi:2014,Cao.Lu:15,Tang.Qu:15}.  The main idea of such methods is to discover latent
dependency among the graph vertices and find metric embedding of vertices that
captures those relationships. The majority of these works use training methods
inspired from neural-network language modeling, such as skip-gram with negative
sampling. However, no such work exists for finding feature representation of
node-pair instances for the purpose of link prediction in a dynamic network.

In this work, we propose \alg\footnote{\alg\ stands for \textbf{Link} \textbf{to} \textbf{Vec}tor 
in a \textbf{Dy}namic network. 
The proposed methodologies maps node-pairs (links) in a dynamic network to a vector representation.}, a novel learning method for obtaining a feature
representation of node-pair instances, which is specifically suitable for the
task of link prediction in a dynamic network. \alg~ considers the feature
learning task as an optimal coding problem, such that the optimal code of a
node-pair is the desired feature representation. The learning process can be
considered as a two-step compression-reconstruction step, where the first step
compresses the input representation of a node-pair into a code by a non-linear
transformation, and the second step reconstructs the input representation from
the code by a reverse process and the optimal code is the one which yields the
least amount of reconstruction error.  The input representation of a node-pair
is constructed using the connection history and the neighborhood information of
the corresponding nodes (details in Section
\ref{sec:Methodology}). After obtaining an appropriate feature
representation of the node-pairs, a standard supervised learning technique can
be used (we use AdaBoost) for predicting link states at future times in the
given dynamic network.

Below we summarize our contributions in this work:
\begin{itemize}
\item We propose a method (\alg) for finding metric embedding of node-pairs
for the task of link prediction over a dynamic network.

\item We validate the effectiveness of \alg\ node-pair embedding by utilizing it
for link prediction on four real-life dynamic networks.

\item We compare the performance of \alg\ embedding based dynamic link prediction model with multiple
state-of-the-art methods. Our
comparison results show that the proposed method is significantly superior than all
the competing methods.

\end{itemize}

The paper is organized as follows. In Section \ref{sec:relwork} we discuss
related work. Section \ref{sec:problemDefinition} defines the problem. In
Section \ref{sec:Methodology} we discuss the proposed learning method \alg. 
In Section \ref{subsec:SupervisedLinkPredictionModel} we detail the link
prediction method using \alg.
Section \ref{sec:experiemnts} presents the experimental results to
validate the effectiveness of our method. Finally, Section \ref{sec:conclusion}
concludes the paper.


\section{Related Works}\label{sec:relwork}
In recent years, the link prediction problem has been studied using a multitude
of methodologies.  The earliest link prediction methodologies use topological
features in a supervised classification
setting~\cite{Hasan06linkprediction,Liben.Nowell:2007}. More recent
methodologies use matrix factorization based approach~\cite{Menon:2011}. Such
methodologies learn latent node representation and predict link strength by the
dot product of the latent vectors of corresponding nodes. The objective
function of these methods may contain appropriate penalty terms for
regularization, and also terms for explicit node and edge features (if
available).  Recently, Bayesian nonparametric latent feature models have also
been proposed for link prediction~\cite{Miller.Jordan.ea:09}.  Unfortunately,
all the above methods fail to capture the temporal evolution of the network on
a dynamic network setting.

A few methods have been developed for link prediction on dynamic networks. The
method proposed by G{\"u}ne{\c{s}} et al. \cite{Gunes2015} capture temporal
patterns in a dynamic network using a collection of time-series on topological
features.  But this approach fails to capture signals from neighborhood
topology, as each time-series model is trained on a separate $t$-size 
feature sequence of a node-pair. Matrix and tensor
factorization based solutions are presented in \cite{Dunlavy:2011}.
Given a three dimensional tensor
representation of a dynamic network, the proposed methods use CANDECOMP/PARAFAC (CP)
decomposition to capture structural and temporal patterns in the dynamic
network. We observe that these methods work well for smaller network, but their 
prediction performance becomes worse as the network grow larger.

The nonparametric link prediction method presented in \cite{ICML2012Sarkar_828} uses features of the node-pairs, as well as the local neighborhood of node-pairs. This method works by choosing a probabilistic model based on features (common neighbor and last time of linkage) of node-pairs.
Stochastic block model based approaches
\cite{KevinS:2013,KevinS:2015} divide nodes in a network into
several groups and generates edges with probabilities dependent on the group
membership of participant nodes. 
%
While
probabilistic model based link prediction performs well on small networks, they become
computationally prohibitive for large networks. 
A deep learning based solution proposed by Li et al. \cite{Li:2014} uses a collection of Restricted Boltzmann Machines with neighbor influence for link prediction in dynamic networks. Tylenda et al. \cite{Tylenda:2009} proposed time-aware link prediction method for evolving social networks with hyper-edges.
 

\section{Problem Definition}\label{sec:problemDefinition}
Let $G(V,E)$ be an undirected network, where $V$ is the set of nodes and $E$
is the set of edges $e(u,v)$ such that $u,v \in V$. A dynamic network is
represented as a sequence of snapshots $\mathbb{G} = \{G_1,G_2, \dots ,G_t\}$,
where $t$ is the number of time stamps for which we have network snapshots and
$G_i(V_i,E_i)$ is a network snapshot at time stamp $i: 1 \le i \le t$.  In this
work, we assume that the vertex set remains the same across different
snapshots, i.e., $V_1 = V_2 = \dots = V_t=V$. However, the edges appear and
disappear over different time stamps. We also assume that, in addition to the link information, no other attribute data for the nodes or edges are available.

Adjacency matrix representation of a network snapshot $G_i$ is represented by a
symmetric binary matrix $\mathbf{A}_i (n \times n)$, where $n$ is the number of
vertices in $G_i$. For two vertices $u$ and $v$,
$\mathbf{A}_i(u,v)=\mathbf{A}_i(v,u)=1$, if an edge exists between them in
$G_i$, and $0$ otherwise. The adjacency vector of a node $u$ at snapshot $G_i$
is a $1 \times n$ row vector defined as $\bm{a}_i^{u} = \mathbf{A}_i(u,1:n)$. \\

\noindent \textbf{Problem Statement:} Given a sequence of snapshots $\mathbb{G}
= \{G_1,G_2, \dots ,G_t\}$ of a network, the task of metric embedding of the
node-pairs $(u,v)$ is to obtain a vector $\bm{\alpha}^{uv} \in \mathbb{R}^l$
($l$ is the dimensionality of embedding) such that node-pairs having similar 
local structures across different time snapshots are packed together in the
embedding. Once such metric embedding of a node-pair $(u,v)$ is
obtained, we use it as the feature representation of this node-pair while predicting the
link status between $u$ and $v$ in $G_{t+1}$. Note that, we assume that no link 
information regarding the snapshot $G_{t+1}$ is available, except the fact that 
$G_{t+1}$ contains the identical set of vertices.

\section{Metric Embedding of Node-Pairs}\label{sec:Methodology}
A key challenge for dynamic link prediction is choosing an effective metric
embedding for a given node-pair. Earlier works construct feature vector by
adapting various topological similarity metrics for static link prediction or
by considering the feature values of different snapshots as a time-series.
\alg, on the other hand, learns the feature embedding of the node-pairs by
using an optimization framework, considering both network topology and link
history. Assume a node-pair $(u,v)$ for which we are computing the metric
embedding $\bm{\alpha}^{uv} \in \mathbb{R}^d$.
Since we want $\bm{\alpha}^{uv}$ to facilitate link prediction on
dynamic graphs, the vector $\bm{\alpha}^{uv}$ must capture two aspects that
influence the possibility of link between $u$ and $v$ in $G_{t+1}$. The first
aspect is the similarity between $u$ and $v$ in terms of graph topology across
different timestamps, and the second aspect is the history of collaboration
between $u$ and $v$---both in the graph snapshots $G_1, \cdots, G_t$.

Consideration of first aspect requires to impart topological similarity signals
between $u$ and $v$ into the desired embedded vector $\bm{\alpha}^{uv}$ by
considering $u$ and $v$'s relation across all the timestamps.  To fulfill this
objective, we start with a feature vector, $\bm{a}^{uv}_{[1,t]}$
of size $nt$ for a node pair $(u,v)$
by taking the element-wise summation of adjacency vectors of $u$ and $v$ over
all the timestamps.  Thus, for a snapshot $G_i$, the adjacency summation vector
is $\bm{a}_i^{uv}=\bm{a}_i^{u}+\bm{a}_i^{v}$, and the entire feature vector is
the concatenation of $\bm{a}_i^{uv}$'s from a continuous set of network
snapshots, i.e., $\bm{a}^{uv}_{[1,t]}
=\bm{a}_1^{uv}~||~\bm{a}_2^{uv}~||~\dots~||~\bm{a}_{t}^{uv}$.  Here, the symbol
$||$ represents concatenation of two horizontal vectors ( e.g., $0~1~0 ~||
~0.5~0~1 = 0~1~0~0.5~0~1$ ). \\

\begin{figure}[t]
\centering
\includegraphics[scale=.4]{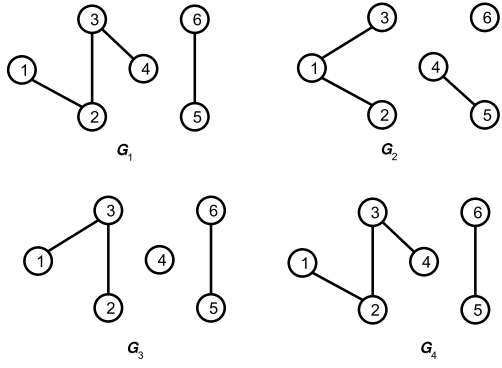}
\caption{A toy dynamic network $\mathbb{G}$ with four snapshots $G_1$, $G_2$ , $G_3$ and $G_4$. Note that the number of nodes remains constant ($6$) even though the links (edges) may change over time.}
\label{fig:example}
\end{figure}

\begin{figure}[t]
\centering
\includegraphics[]{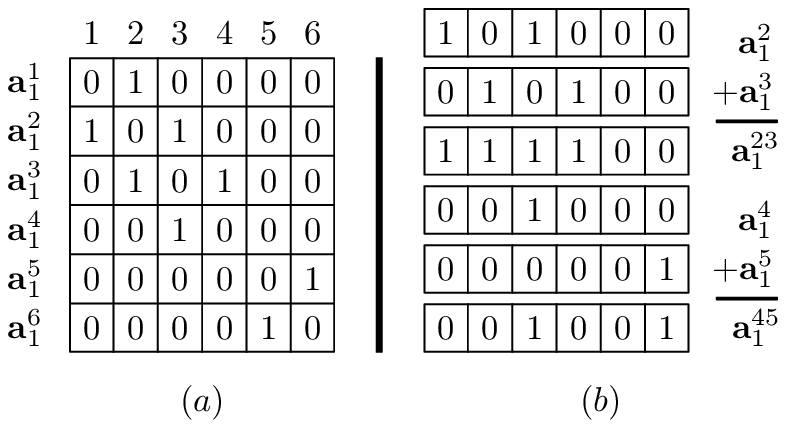}
\caption{(a) Adjacency matrix $\bm{A}_1$. Each row represents adjacency vector of the corresponding node (b) Computation of node-pair adjacency vector $\bm{a}_1^{23}$ and $\bm{a}_1^{45}$. }
\label{fig:basicFeature}
\vspace{-0.2in}
\end{figure}

{\noindent} {\bf Example:} Consider the toy dynamic network shown in Figure
\ref{fig:example}. The dynamic network $\mathbb{G} = \{G_1,G_2,G_3,G_4\}$ has
four snapshots. The task is to predict the edges in snap $G_5$ (not shown in
this figure). The set of nodes does not change over time $(V_1=V_2=\cdots=V_5)$.
In Figure \ref{fig:basicFeature}(a) we show the adjacency matrix of the
dynamic network $\mathbb{G}$ at time-stamp 1. In
Figure \ref{fig:basicFeature}(b), we show the computation of adjacency
vectors of two node-pairs, namely $\bm{a}_1^{23}$ and $\bm{a}_1^{45}$. $\qed$.

The second aspect, history of collaboration between a node-pair is captured by
taking cumulative sum of link history, weighted by a time decay function.

\begin{equation*}
\mathbf{wclh}^{uv}_{[1,t]}  = CumSum(\mathbf{wlh}^{uv}_{[1,t]})
\end{equation*}
Here 
$\mathbf{wlh}^{uv}_{[1,t]} = w_1\cdot\mathbf{A}_1(u,v)~||~w_2\cdot\mathbf{A}_2(u,v)~||~\dots~
||~w_{t}\cdot\mathbf{A}_{t}(u,v)$ and $w_i=i/t$ is the time decay function. Time decay function $w_i$ prioritize more recent linkage information, while cumulative sum rewards reappearance of links (between $u$ and $v$) over different time snapshots.

Finally, the feature vector for a node-pair ($u,v$),
$\mathbf{e}^{uv}$, is the concatenation of 
$\bm{a}^{uv}_{[1,t]}$ and 
($\mathbf{wclh}^{uv}_{[1,t]}$); i.e.,  $\mathbf{e}^{uv} =
\bm{a}^{uv}_{[1,t]}~||\mathbf{wclh}^{uv}_{[1,t]}$. \alg's optimization
framework converts $\bm{e}^{uv}$ to the optimal feature representation
$\bm{\alpha}^{uv}$ by using a non-linear transformation function $h$ discussed 
in Section \ref{subsec:UnsupervisedFeatureExtraction}. Note that, through
$h$, the proposed method models complex functions of
the entries in $\bm{e}^{uv}$, which makes the embedded feature vector $\bm{\alpha}^{uv}$ 
very effective for link prediction in dynamic network. 

Our proposed method is different---both, in methodologies and also in objective---
from the existing works~\cite{Tang.Qu:15,Perozzi:2014} which construct metric embedding of the vertices of a network. Existing
works find embedding of a vertex from a static network, whereas we find
embedding of a node-pair from a dynamic network. The learning method of the
existing works follow language model, whereas our method follows
a compression-reconstruction framework which preserves
higher-order neighborhood and link history patterns of the node-pair in its
embedded representation. Below we discuss the compression-reconstruction
framework which yields the optimal metric embedding through a principled approach.

\subsection{Optimization framework for DyLink to Vec}\label{subsec:UnsupervisedFeatureExtraction}

In this section, we discuss the optimization framework which obtains the
optimal metric embedding of a node pair by learning an optimal coding function
$h$. For this
learning task, let's assume $\mathbf{\widehat{E}}$ is the training dataset
matrix containing a collection of node-pair feature vectors. Each row of this
matrix represents a node-pair (say, $u$ and $v$) and it contains the feature
vector $\bm{e}^{uv}$ which stores information about neighborhood and link
history, as we discussed earlier. The actual link status of the node-pairs in $\mathbf{\widehat{E}}$ 
in $G_{t+1}$ is not used for the learning of $h$, so the metric embedding process
is unsupervised. In subsequent discussion, we write $\bm{e}$ to represent
an arbitrary node pair vector in $\mathbf{\widehat{E}}$.

Now, the coding function $h$ compresses $\bm{e}$ to a code vector $\bm{\alpha}$
of dimension $l$, such that $l < k$. Here $l$ is a user-defined parameter which
represents the code length and $k$ is the size of feature vector. Many
different coding functions exist in the dimensionality reduction literature,
but for \alg\ we choose the coding function which incurs the minimum
reconstruction error in the sense that from the code $\bm{\alpha}$ we can
reconstruct $\bm{e}$ with the minimum error over all $\bm{e} \in
\mathbf{\widehat{E}}$.  We frame the learning of $h$ as an optimization
problem, which we discuss below through two operations: Compression  and
Reconstruction. \\

\noindent {\bf Compression:} It obtains $\bm{\alpha}$ from $\bm{e}$.
This transformation can be expressed as a nonlinear function of linear weighted
sum of the entries in vector $\bm{e}$.
\begin{equation}
\bm{\alpha} = f(\bm{W}^{(c)}\bm{e} + \bm{b}^{(c)})
\label{eq:Compression}
\end{equation}
$\bm{W}^{(c)}$ is a $(k \times l)$ dimensional matrix. It represents the weight matrix for compression
and $\bm{b}^{(c)}$ represents biases. $f(\cdot)$ is the Sigmoid function, $f(x)=\frac{1}{1+e^{-x}}$. \\

\noindent {\bf Reconstruction:} It performs the reverse operation of compression, i.e., it 
obtains $\mathbf{e}$ from $\bm{\alpha}$ (which was constructed during the compression 
operation).
\begin{equation}
\bm{\beta} = f(\bm{W}^{(r)}\bm{\alpha} + \bm{b}^{(r)})
\label{eq:Reconstruction}
\end{equation}
$\bm{W}^{(r)}$ is a matrix of dimensions $(l \times k)$ representing the weight matrix for reconstruction, 
and $\bm{b}^{(r)}$ represents biases. 

The optimal coding function $h$ constituted by the compression and reconstruction operations is defined by the parameters  $(\bm{W},\bm{b}) = (\bm{W}^{(c)},\bm{b}^{(c)},\bm{W}^{(r)},\bm{b}^{(r)})$.
The objective is to minimize the reconstruction error.
Reconstruction error for a neighborhood based feature vector $\bm(e)$ is defined as,
$J(\bm{W,b,e}) = \frac{1}{2}\parallel \bm{\beta}-\bm{e} \parallel^2$.
Over all possible feature vectors, the average reconstruction error augmented with a regularization
term yields the final objective function $J(\bm{W},\bm{b})$:
\begin{equation}
\begin{split}
J(\bm{W},\bm{b}) = \frac{1}{{|\mathbf{\widehat{E}}|}} \sum_{\bm{e} \in \mathbf{\widehat{E}}}(\frac{1}{2}\parallel \bm{\beta}^{uv}-\bm{e}^{uv} \parallel^2)
\\+\frac{\lambda}{2} (\parallel\bm{W}^{(c)}\parallel_F^2 +  \parallel\bm{W}^{(r)}\parallel_F^2 )
\end{split}
\label{eq:lossfun2}
\end{equation}

Here, $\lambda$ is a user assigned regularization parameter, responsible for
preventing over-fitting. $\parallel\cdot\parallel_F$ represents the 
Frobenius norm of a matrix. In this work we use $\lambda = 0.1$. 

To this end, we discuss the motivation of our proposed optimization framework
for learning the coding function $h$. Note that,
the dimensionality of $\bm{\alpha}$ is much smaller than $\bm{e}$, so the
optimal compression of the vector $\bm{e}$ must extract patterns composing of
the entries of $\bm{e}$ and use them as high-order latent feature in
$\bm{\alpha}$. In fact, the entries in $\bm{e}$ contain the neighborhood (sum
of adjacency vector of the node pair) and link history of a node-pair for all
the timestamps; for a real-life network, this vector is sparse and substantial
compression is possible incurring small loss. Through this compression the
coding function $h$ learns the patterns that are similar across different node-pairs (used in $\mathbf{\widehat{E}}$).
Thus the function $h$ learns a metric embedding of the node-pairs that packs node-pairs
having similar local structures in close proximity in the embedded feature space.
Although function $h$ acts as a black-box, it captures patterns involving neighborhood
around a node pair across various time stamps, which obviates the manual construction 
of a node-pair feature---a cumbersome task for the case of a dynamic network.

\subsubsection{Optimization}

The training of optimal coding defined by parameters $(\bm{W,b})$ begins with
random initialization of the parameters. Since the cost function
$J(\bm{W},\bm{b})$ defined in Equation \eqref{eq:lossfun2} is non-convex in
nature, we obtain a local optimal solution using the gradient descent approach.
Such approach usually provides practically useful results (as shown in the Section 
\ref{sec:experiemnts}).
The parameter updates of the gradient descent are similar to the parameter updates
for optimizing Auto-encoder in machine learning.  
One iteration of gradient descent updates the parameters using following equations:

\begin{equation}
\begin{split}
W_{ij}^{(c)} = W_{ij}^{(c)} - \sigma \frac{\partial}{\partial W_{ij}^{(c)}}J(W,b) \\
W_{ij}^{(r)} = W_{ij}^{(r)} - \sigma \frac{\partial}{\partial W_{ij}^{(r)}}J(W,b) \\
b_{i}^{(c)} = b_{i}^{(c)} - \sigma \frac{\partial}{\partial b_{i}^{(c)}}J(W,b) \\
b_{i}^{(r)} = b_{i}^{(r)} - \sigma \frac{\partial}{\partial b_{i}^{(r)}}J(W,b) 
\end{split}
\label{eq:updatefun1}
\end{equation}

Here, $l$ appropriately identifies the weight and bias parameters $l \in  \{1,2\}$. $\sigma$ is the learning rate. $W_{ij}^{(1)}$ is the weight of connection between node $j$ of the input layer to node $i$ of the hidden layer. 

Now, from Equation \eqref{eq:lossfun2}, the partial derivative terms in equations \eqref{eq:updatefun1} can be written as,
\begin{equation}
\begin{split}
\frac{\partial}{\partial W_{ij}^{(c)}}J(W,b) =  \frac{1}{|\mathbf{\widehat{E}}| } \sum_{\bm{e} \in \mathbf{\widehat{E}}}\frac{\partial}{\partial W_{ij}^{(c)}}J(\mathbf{W,b,e})+\lambda W_{ij}^{(c)} \\
\frac{\partial}{\partial W_{ij}^{(r)}}J(W,b) =  \frac{1}{|\mathbf{\widehat{E}}|} \sum_{\bm{e} \in \mathbf{\widehat{E}}}\frac{\partial}{\partial W_{ij}^{(r)}}J(\mathbf{W,b,e})+\lambda W_{ij}^{(r)} \\
\frac{\partial}{\partial b_{i}^{(c)}}J(W,b) =  \frac{1}{|\mathbf{\widehat{E}}| } \sum_{\bm{e} \in \mathbf{\widehat{E}}}\frac{\partial}{\partial b_{i}^{(c)}}J(\mathbf{W,b,e}) \\
\frac{\partial}{\partial b_{i}^{(r)}}J(W,b) =  \frac{1}{|\mathbf{\widehat{E}}| } \sum_{\bm{e} \in \mathbf{\widehat{E}}}\frac{\partial}{\partial b_{i}^{(r)}}J(\mathbf{W,b,e})  \\
\end{split}
\label{eq:derivative}
\end{equation}

The optimization problem is solved by computing partial derivative of cost function $J(\mathbf{W,b,e})$ using the back propagation approach \cite{Rumelhart:1986}. 

Once the optimization is done,
the  metric embedding of
any node-pair ($u,v$) can be obtained by taking the outputs of compression
stage (Equation \eqref{eq:Compression}) of the trained optimal coding
($\bm{W,b}$).

\begin{equation}
\begin{split}
\bm{\alpha}^{uv} = f(\bm{W}^{(c)}\bm{e}^{uv} + \bm{b}^{(c)})=h(\bm{e}^{uv})
\end{split}
\label{eq:unsupervisedFeatures}
\end{equation}

\subsubsection{Complexity Analysis}
We use Matlab implementation of optimization algorithm L-BFGS (Limited-memory
Broyden-Fletcher-Goldfarb-Shanno) for learning optimal coding.  We execute the
algorithm for a limited number of iterations to obtain unsupervised features
within a reasonable period of time. Each iteration of L-BFGS executes two tasks
for each node-pair: back-propagation to compute partial differentiation of cost
function, and change the parameters $\bm{(W,b)}$. 
Therefore, the time complexity
of one iteration is $O(|NP_t|kl)$. Here, $NP_t$ is the set on node-pairs used to construct the training dataset $\mathbf{\widehat{E}}$.
$k$ is the length of $\bm{e}$ (dimensionality of initial edge features), and $l$ is
length of $\bm{\alpha}$ (optimal coding).

\section{Link Prediction using proposed metric embedding}\label{subsec:SupervisedLinkPredictionModel}
For link prediction task in a dynamic network, $\mathbb{G} = \{G_1,G_2, \dots ,G_t\}$; we split the snapshots into two overlapping time windows, $[1,t-1]$ and $[2,t]$.
Training dataset, $\mathbf{\widehat{E}}$ is feature representation for time
snapshots $[1,t-1]$, the ground truth ($\widehat{\bm{y}}$) is constructed from
$G_t$. \alg\ learns optimal embedding $h(\cdot)$ using training dataset $\mathbf{\widehat{E}}$.
After training a supervised classification model using
$\widehat{\bm{\alpha}}$=$h(\widehat{\mathbf{E}})$ and $\widehat{\bm{y}}$, prediction
dataset $\mathbf{\overline{E}}$ is used to predict links at $G_{t+1}$. For this
supervised prediction task, we experiment with several classification
algorithms. Among them SVM (support vector machine) and
AdaBoost perform the best.

\begin{algorithm}[h]
	\SetKwInOut{Input}{Input}
    \SetKwInOut{Output}{Output}
\begin{algorithmic}[1]
		\Procedure{LP\alg}{$\mathbb{G},t$}
		\Input{$\mathbb{G}$: Dynamic Network, $t$: Time steps}
    		\Output{$\overline{y}$: Forecasted links at time step $t+1$}
    		
		\State $\widehat{\mathbf{E}}$=NeighborhoodFeature($\mathbb{G}$,$1$,$t-1$)
		\State $\widehat{\mathbf{y}}$=Connectivity($G_t$)
		\State $\overline{\mathbf{E}}$=NeighborhoodFeature($\mathbb{G}$,$2$,$t$)
		\State $h$=LearningOptimalCoding($\widehat{\mathbf{E}}$)
		\State $\widehat{\bm{\alpha}}$=$h(\widehat{\mathbf{E}})$
		\State $\overline{\bm{\alpha}}$=$h(\overline{\mathbf{E}})$
		\State $C$=TrainClassifier($\widehat{\bm{\alpha}},\widehat{\mathbf{y}}$)
		\State $\overline{\mathbf{y}}$=LinkForecasting($C,\overline{\bm{\alpha}}$)
		\State \textbf{return} $\overline{\mathbf{y}}$
		\EndProcedure
		\end{algorithmic}
		\caption{Link Prediction using \alg}\label{alg:LPUFE}
\end{algorithm}

The pseudo-code of \alg\ based link prediction method is given in Algorithm \ref{alg:LPUFE}.  For training
link prediction model, we split the available network snapshots into two
overlapping time windows, $[1,t-1]$ and $[2,t]$. Neighborhood based features
$\widehat{\bm{E}}$ and $\overline{\bm{E}}$ are constructed in Lines 2 and 4,
respectively. Then we learn optimal coding for node-pairs using neighborhood
features $\widehat{\bm{E}}$ (in Line 5). Embeddings
are constructed using learned optimal coding (Lines 6 and 7) using output of
compression stage (Equation \ref{eq:unsupervisedFeatures}). Finally, a
classification model $C$ is learned (Line 8), which is used for predicting
links in $G_{t+1}$ (Line 9).

\section{Experiments and Results}\label{sec:experiemnts}
We demonstrate the performance of \alg\ using four real world
dynamic network datasets: {\bf Enron}, {\bf Collaboration}, {\bf Facebook1} and
{\bf Facebook2}. We show performance comparison between \alg\ based link prediction method and existing
state-of-the-art dynamic link prediction methodologies.  Experimental results
also include the discussion of \alg's performance for varying length of time
stamps in the network, and varying degree of class imbalance in training dataset.  Bellow, we discuss
the datasets, evaluation metrics, competing methods, implementation details and
results.

\begin{figure}[!hp] 
\subfloat[Enron Network]{%
\includegraphics[width=.43\linewidth]{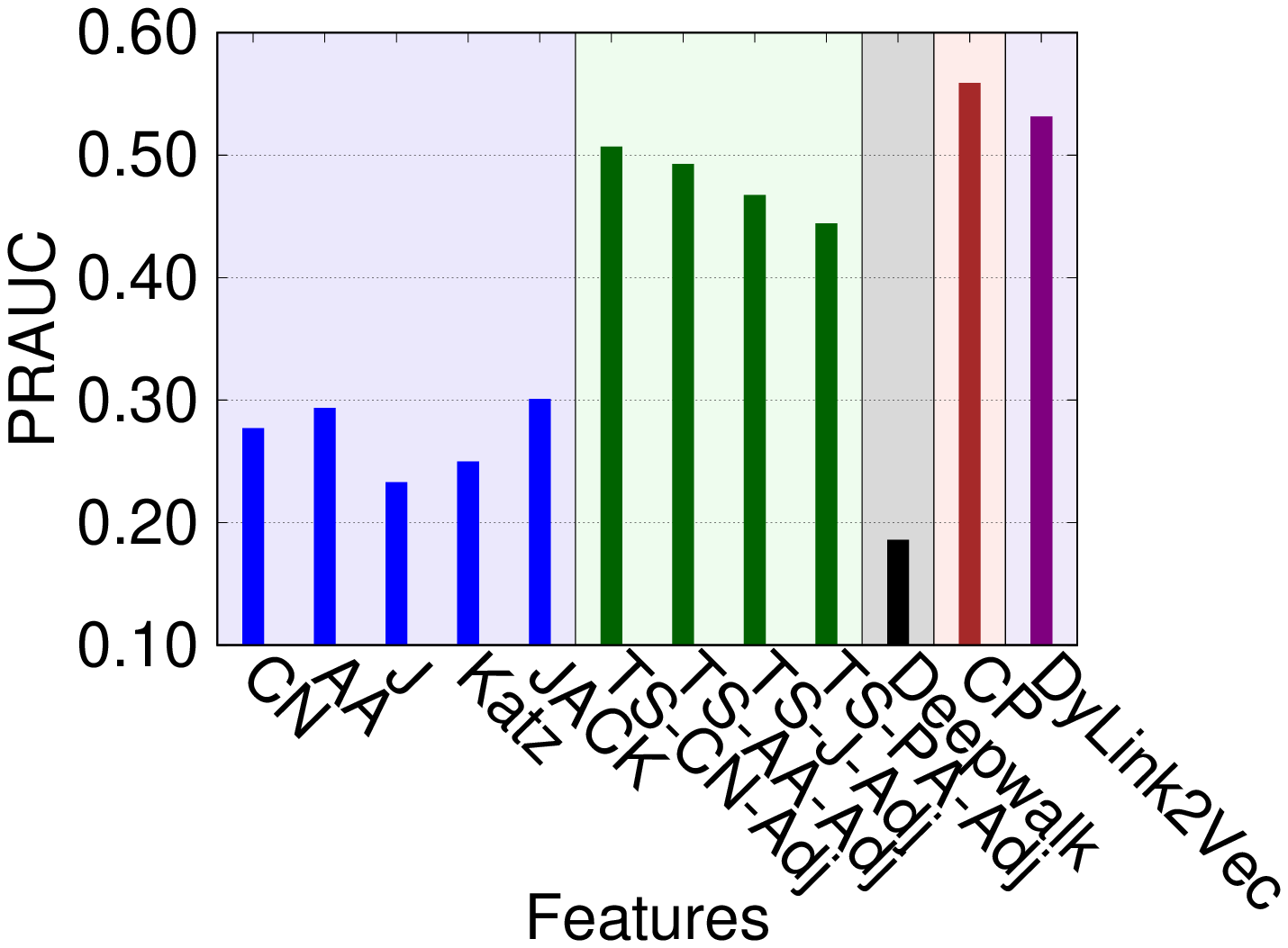}
} 
\subfloat[Enron Network]{%
\includegraphics[width=.43\linewidth]{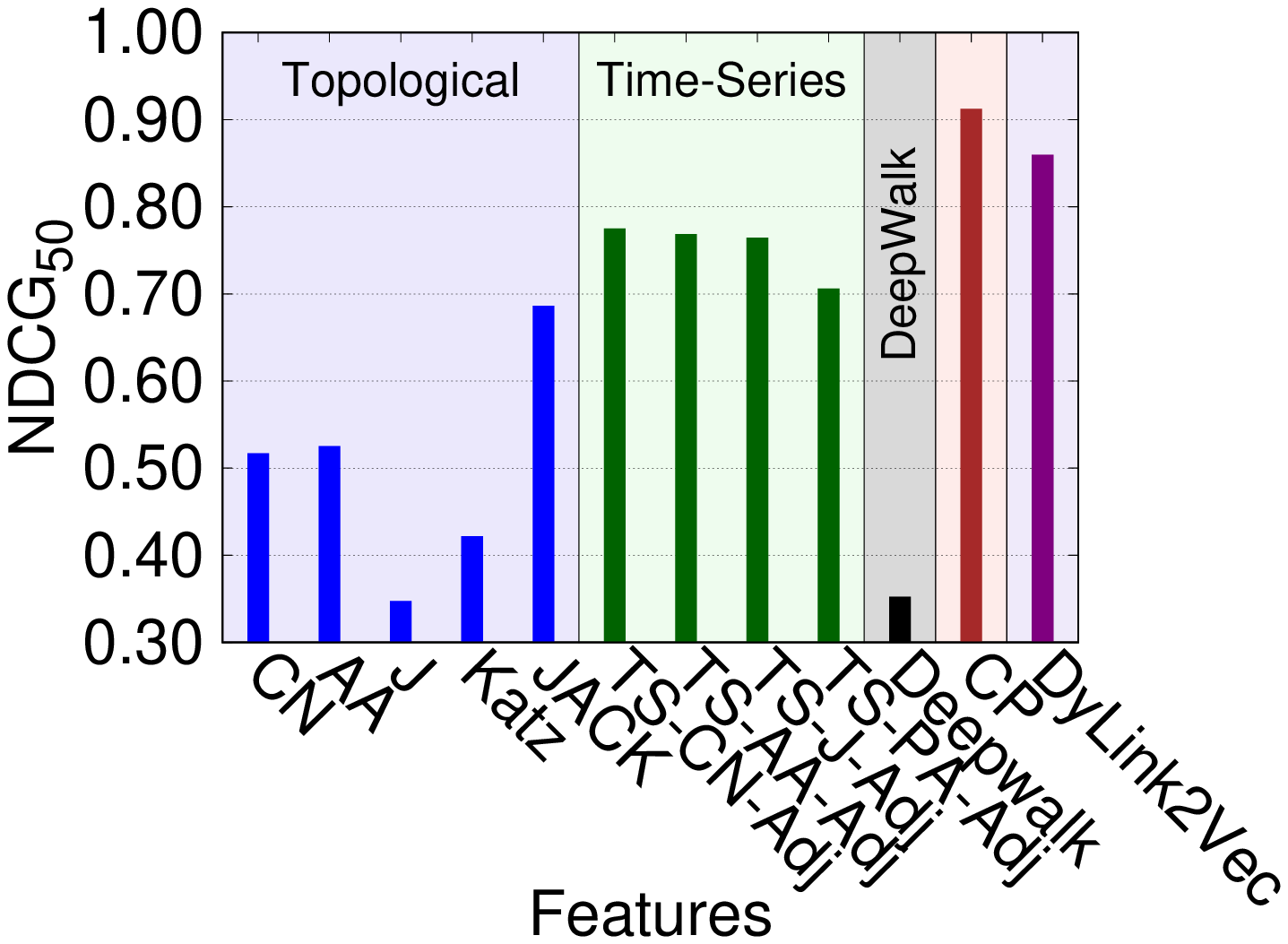}
}
\\
\subfloat[Collaboration Network]{%
\includegraphics[width=.43\linewidth]{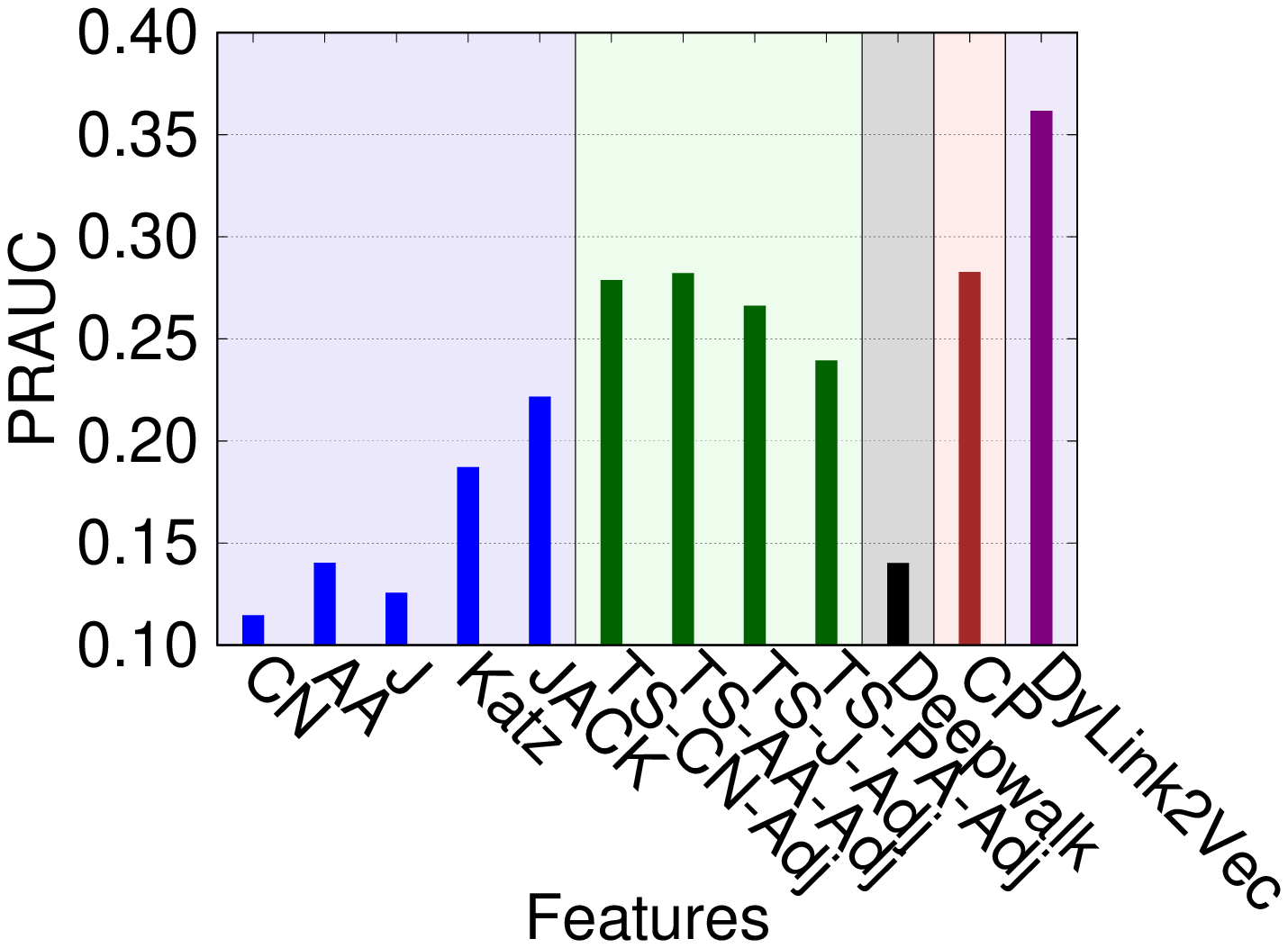}
} 
\subfloat[Collaboration Network]{%
\includegraphics[width=.43\linewidth]{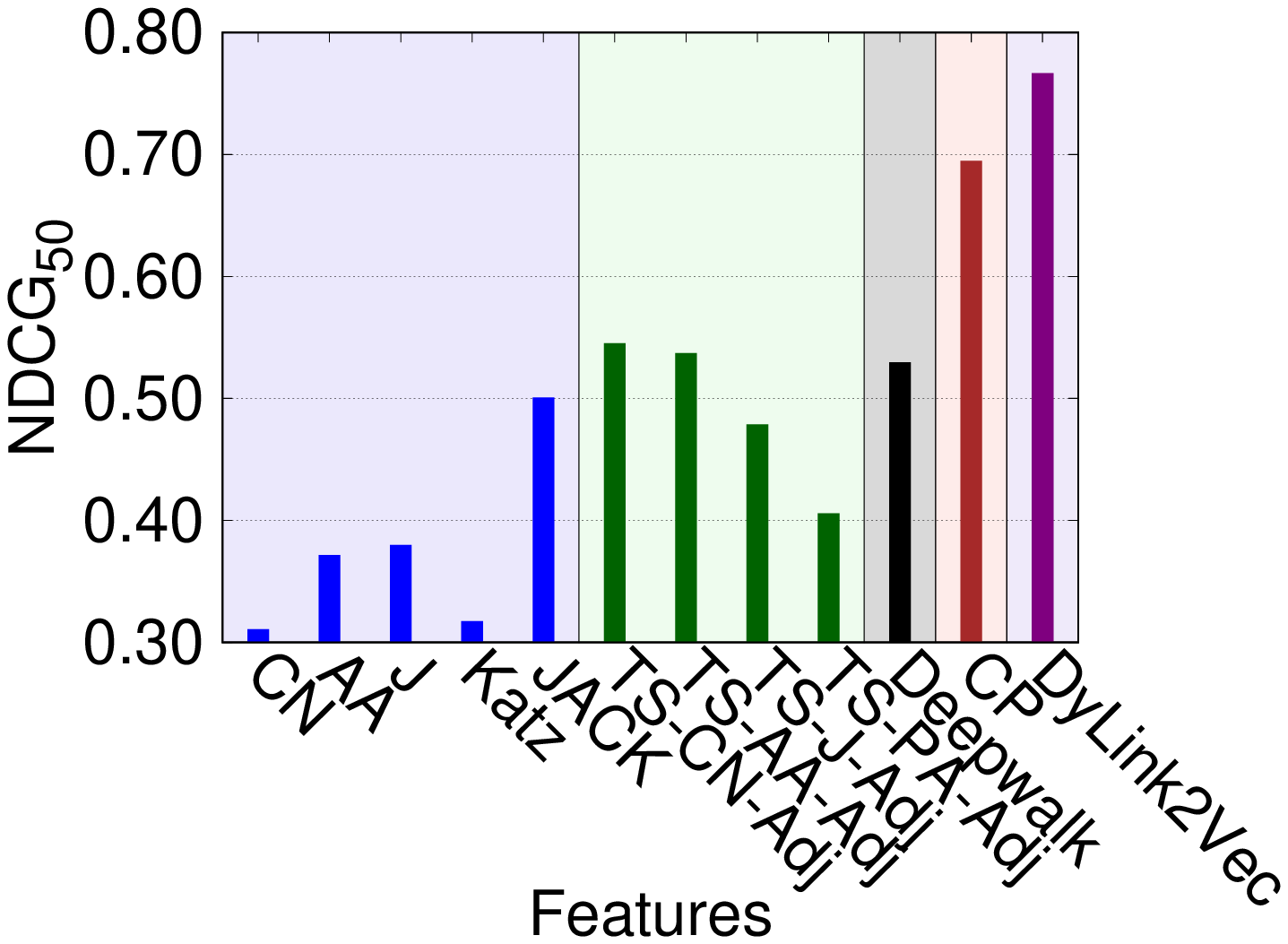}
}
\\ 

\subfloat[Facebook1 Network]{%
\includegraphics[width=.43\linewidth]{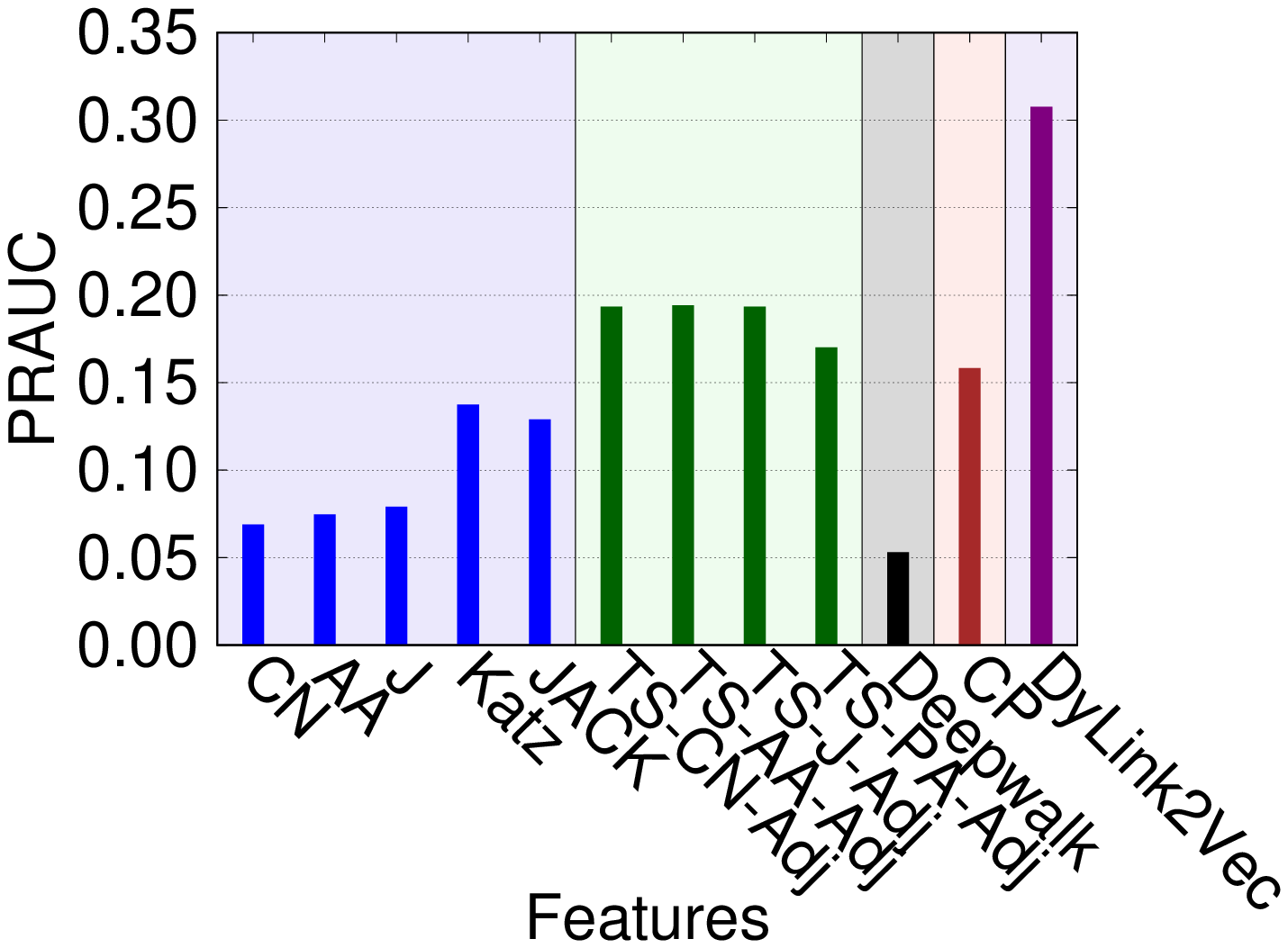}
} 
\subfloat[Facebook1 Network]{%
\includegraphics[width=.43\linewidth]{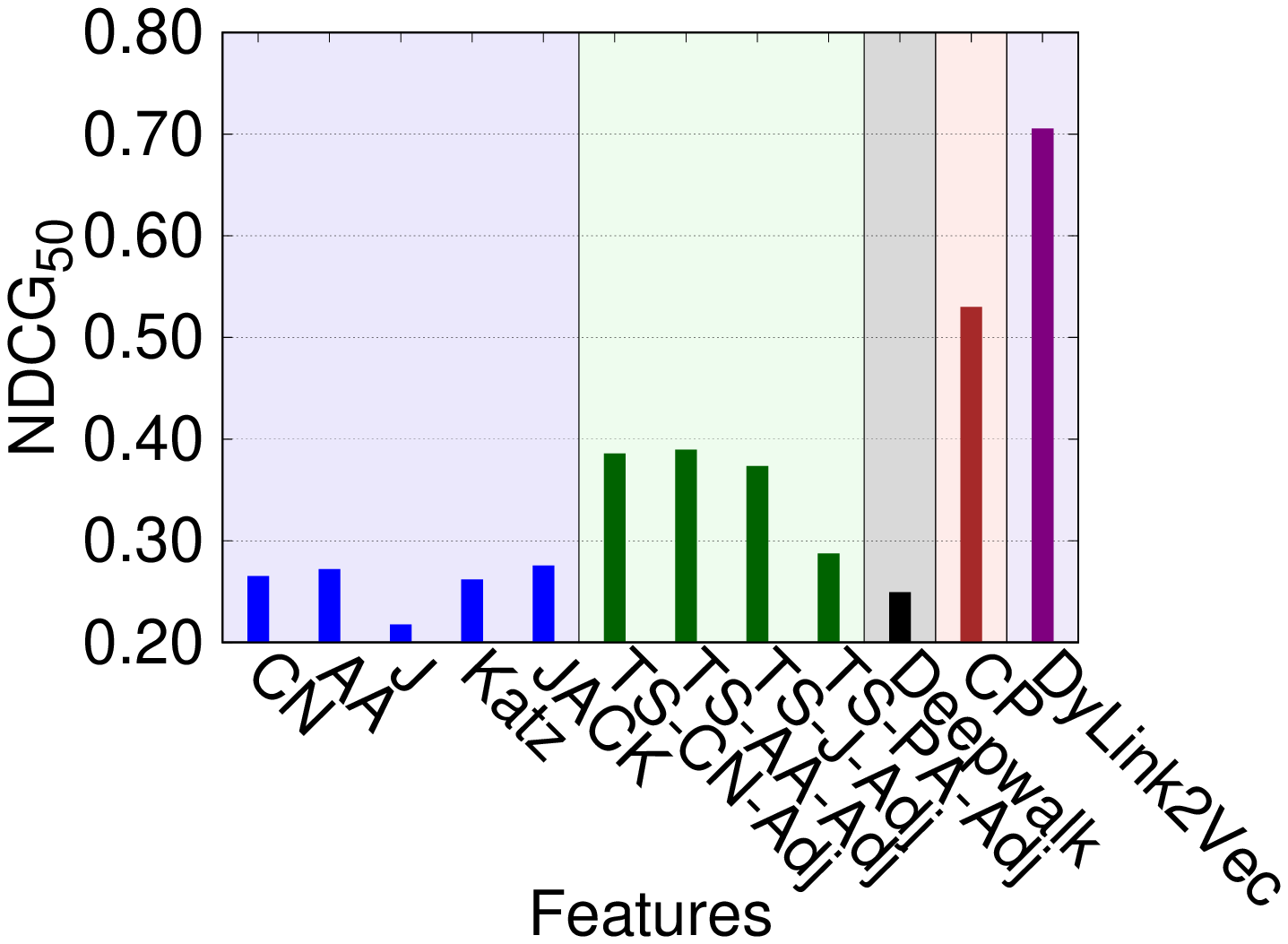}
}
\\
\subfloat[Facebook2 Network]{%
\includegraphics[width=.43\linewidth]{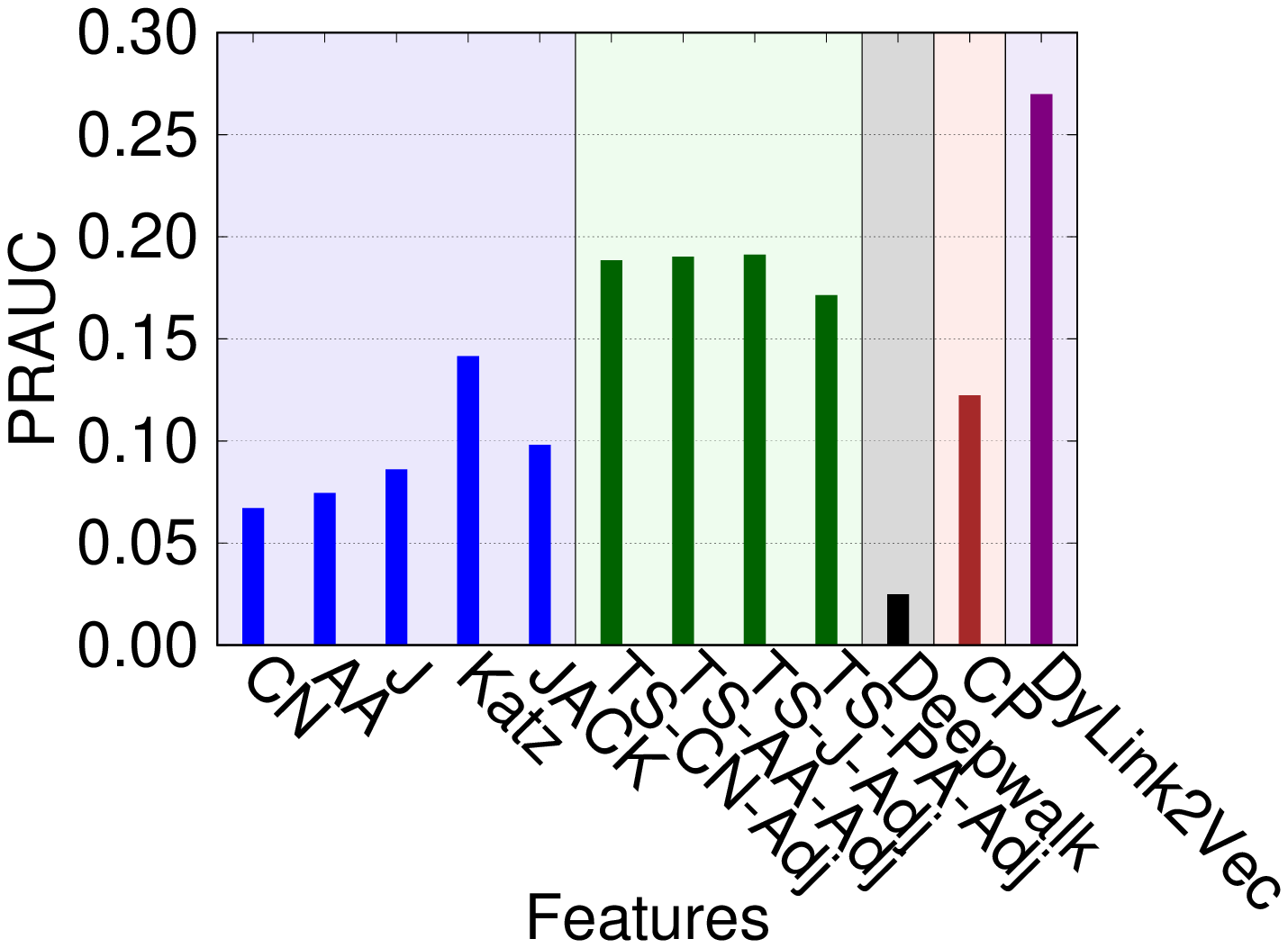}
} 
\subfloat[Facebook2 Network]{%
\includegraphics[width=.43\linewidth]{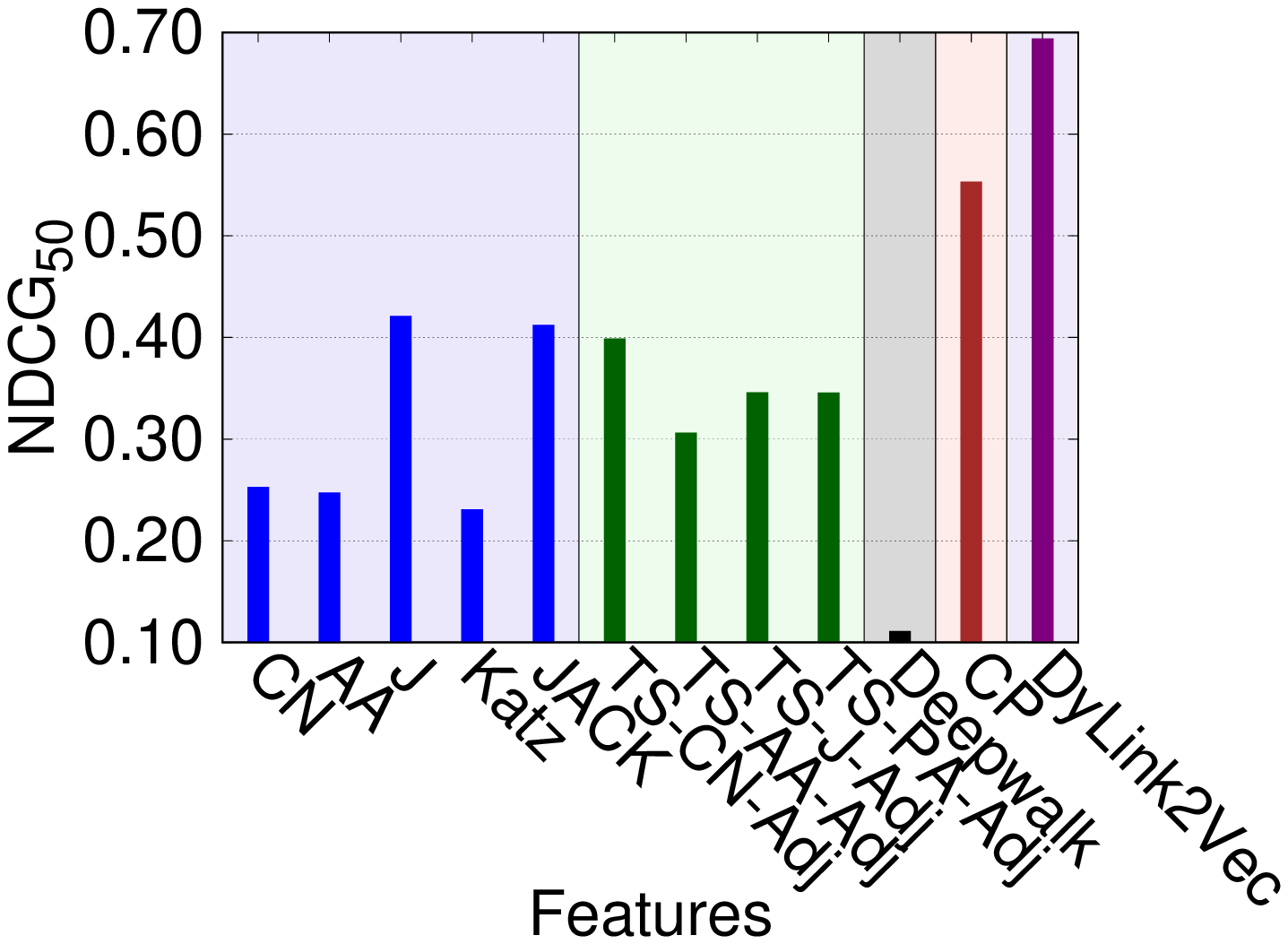}
}
\caption{Comparison with competing link prediction methods. Each bar
represents a methods and the height of the bar represents the value of the
performance metrics. Results for Enron network are presented in charts (a)-(b), results of Collaboration data are presented in charts (c)-(d), the results of Facebook1 data are in charts (e)-(f), and the results of Facebook2 data are in charts (g)-(h). The group of bars in a chart are distinguished by
color, so the figure is best viewed on a computer screen or color print.}
\label{fig:CompareAll} 
\end{figure}

\subsection{Dataset Descriptions}\label{sec:DataSet}

Here we discuss the construction and characteristics of the datasets used for
experiments. 

\noindent {\em  Enron} email corpus \cite{priebe2005scan} consists of email
exchanges between Enron employees. The Enron dataset has $11$ time
stamps and $16,836$ possible node-pairs; the task is to use first $10$ snapshots for predicting links in the
$11^{th}$ snapshot. Following \cite{KevinS:2013}, we aggregate data into time
steps of $1$ week. We use the data from weeks $147$ to $157$ of the data trace
for the  experiments. The reason for choosing that window is that the snapshot
of the graph at week $157$ has the highest number of edges. \\

{\noindent} {\em Collaboration} dataset has $10$ time stamps with author collaboration information about $49,455$ author-pairs. The Collaboration dataset is constructed from citation
data containing $1.4$ million papers \cite{Tang:2008}. We process the data to
construct a network of authors with edges between them if they co-author a
paper. Considering each year as a time stamp, the data of years $2000$-$2009$
($10$ time stamps) is used for this experiment, where the data from the first
nine time stamps is used for training and the last for prediction.  Since this
data is very sparse, we pre-process the data to retain only the active authors,
who have last published papers on or after year $2010$; moreover, the selected
authors participate in at least two edges in seven or more time stamps.  \\

{\noindent} {\em  Facebook1 and Facebook2} are network of Facebook 
wall posts \cite{viswanath2009evolution}. Each vertex is a Facebook user account and
an edge represents the event that one user posts a message on the wall of another user.
Both Facebook1  and Facebook2 has $9$ time stamps. Facebook1 has
$219,453$ node-pairs. Facebook2 is an extended version of Facebook1 dataset with $883,785$
node-pairs. For pre-processing Facebook1 we follow the same setup as is discussed in
\cite{KevinS:2015}; wall posts of 90 days are aggregated in one time step.

We filter out all people who are active for less than 6 of the 9 
time steps, along with the people who have degree less than 30. 
Facebook2 is created using a similar
method, but a larger sample of Facebook wall posts is used for this dataset.

\subsection{Evaluation Metrics}
For evaluating the proposed method we use two metrics, namely, area under
Precision-Recall (PR) curve (PRAUC) \cite{Davis:2006} and an information
retrieval metric, Normalized Discounted Cumulative Gain (NDCG). PRAUC is best suited for evaluating two class classification performance when class
membership is skewed towards one of the classes. This is exactly the case for
link prediction; the number of edges $(|E|)$ is very small compared to the
number of possible node-pairs ${|V|\choose 2}$  In such scenarios, area under the Precision-Recall curve (PRAUC) gives a more informative assessment of the algorithm's performance than other metrics such as, accuracy.  The reason why PRAUC is more suitable for the skewed problem is that it does not factor in  the count of true negatives in its calculation. In skewed data where the number of negative
examples is huge compared to the number of positive examples, true negatives
are not that meaningful. 

We also use NDCG, an information
retrieval metric (widely used by the recommender systems community) to evaluate
the proposed method.  NDCG measures the performance of link prediction system
based on the graded relevance of the recommended links. $NDCG_k$ varies from
0.0 to 1.0, with 1.0 representing ideal ranking of edges. Here, $k$ is a parameter
chosen by user representing the number of links ranked by the method. We use
$k=50$ in all our experiments.

Some of the earlier works on link prediction have used area under the ROC curve
(AUC) to evaluate link prediction works \cite{Gunes2015,Wang:2007}. But recent
works \cite{Yang2015} have demonstrated the limitations of AUC and argued
in favor of PRAUC over AUC for evaluation of link prediction. So we have not used AUC in this work.

\subsection{Competing Methods for Comparison}

We compare the performance of \alg\ based link prediction method with methods from four categories: (1) topological feature based methods, (2) feature time series
based methods~\cite{Gunes2015}, (3) a deep learning based method, namely
DeepWalk~\cite{Perozzi:2014}, and (4) a tensor factorization based method
CANDECOMP/PARAFAC (CP)~\cite{Dunlavy:2011}. 

Besides these four works, there are two other existing works for link
prediction in dynamic network setting; one is based on deep Learning
\cite{Li:2014} (Conditional Temporal Restricted Boltzmann machine) and the other is based on a signature-based nonparametric method \cite{ICML2012Sarkar_828}. We did not compare with these models as implementations of their models are not readily available, besides, both of these methods have numerous parameters which will make reproducibility of their results highly improbable and thus, conclusion derived from such experiments may not align with true understanding of the usefulness of the methods. Moreover, none of these methods give unsupervised feature representation for node-pairs in which we claim our main contribution. 


For \textbf{topological feature based methods}, we consider four prominent topological
features: Common Neighbors ($CN$), Adamic-Adar ($AA$), Jaccard's
Coefficient ($J$) and Katz measure ($Katz$). 
However, in existing works, these features are defined for
static networks only; so we adapt these features for the dynamic network setting
by computing the feature values over the collapsed\footnote{Collapsed network is
constructed by superimposing all network snapshots(see Figure \ref{fig:StaticLimitaion}).} dynamic network. 

We also combine the above four features to construct a combined
feature vector of length four (\textbf{J}accard's
Coefficient, \textbf{A}damic-Adar, \textbf{C}ommon Neighbors and \textbf{K}atz), which we call $JACK$ and use it with a classifier to build a
supervised link prediction method, and include this model in our comparison.

Second, we compare \alg\ with \textbf{time-series based} neighborhood similarity scores proposed in \cite{Gunes2015}. In this work, the authors consider several
neighborhood-based node similarity scores combined with connectivity
information (historical edge information). Authors use time-series of
similarities to model the change of node similarities over time. Among $16$
proposed methods, we consider $4$ that are relevant to the link prediction task
on unweighted networks and also have the best performance. 
$TS\mbox{-}CN\mbox{-}Adj$ represents time-series on normalized score of Common
Neighbors and connectivity values at time stamps $[1,t]$.  Similarly, we get
time-series based scores for Adamic-Adar ($TS\mbox{-}AA\mbox{-}Adj$), Jaccard's
Coefficient ($TS\mbox{-}J\mbox{-}Adj$) and Preferential Attachment
($TS\mbox{-}PA\mbox{-}Adj$). 

Third, we compare \alg\ with \textbf{DeepWalk} \cite{Perozzi:2014}, a latent node
representation based method.  We use DeepWalk to construct latent
representation of nodes from the collapsed dynamic
network. Then we construct latent representation of node-pairs by computing
cross product of latent representation of the participating nodes. For example,
if the node representations in a network are vectors of size $l$, then the
representation of a node-pair $(u,v)$ will be of size $l^2$, constructed from
the cross product of $u$ and $v$'s representation. The DeepWalk based
node-pair representation is then used with a classifier to build a supervised
link prediction method. We choose node representation size $l=2,4,6,8,10$ and 
report the best performance.

Finally, we compare \alg\ with a tensor factorization based method, called
\textbf{CANDECOMP/PARAFAC (CP)} \cite{Dunlavy:2011}. In this method,
the dynamic network is represented as a three-dimensional tensor 
$\mathcal{Z}(n\times n\times t)$. Using CP decomposition $\mathcal{Z}$ is factorized
into three factor matrices. The link prediction score is computed by using 
the factor matrices.
We adapted the CP link prediction method for unipartite networks; which has 
originally been developed for bipartite networks.

\subsection{Implementation Details}

We implemented \alg\ algorithm in Matlab version $R2014b$.  The learning method runs for a maximum of $100$ iterations
or until it converges to a local optimal solution.  We use coding size $l=100$
for all datasets\footnote{We experiment with different coding sizes
ranging from 100 to 800. The change in link prediction performance is not
sensitive to the coding size. At most 2.9\% change in PRAUC was observed for
different coding sizes.}.  For supervised link prediction step we use several
Matlab provided classification algorithms, namely, AdaBoostM1, RobustBoost, and
Support Vector Machine (SVM).  We could use neural network classifier. But, as our main goal is to evaluate the quality of unsupervised feature representation, so, we use simple classifiers. Supervised neural network architecture may result in superior performance, but, it is out of scope of the main goal of the paper. We use Matlab for computing the feature values (CN, AA, J, Katz) that we use in other competing methods. Time-series methods are implemented using Python. We use the ARIMA (autoregressive integrated
moving average) time series model implemented in Python module
\textbf{statsmodels}.  The DeepWalk implementation is provided by the authors
of~\cite{Perozzi:2014}. We use it to extract node features and  extend it for
link prediction (using Matlab). Tensor factorization based method CP was
implemented using Matlab Tensor Toolbox. 

\subsection{Performance Comparison Results with Competing Methods}
In Figure \ref{fig:CompareAll} we present the performance comparison results
of \alg\ based link prediction method with the four kinds of competing methods that we have discussed
earlier. The figure have eight bar charts.  The bar charts from the top to the
bottom rows display the results for Enron, Collaboration, Facebook1 and
Facebook2 datasets, respectively.  The bar charts in a row show comparison
results using PRAUC (left), and  $NDCG_{50}$ (right) metrics.  
Each chart has twelve bars, each representing a link prediction method, where the height
of a bar is indicative of the performance metric value of the corresponding
method. In each chart, from left to right, the first five bars (blue) correspond to the
topological feature based methods, the next four (green) represent time series
based methods, the tenth bar (black) is for DeepWalk, the eleventh bar (brown)
represents tensor factorization based method CP, and the final bar (purple)
represents the proposed method \alg.

\subsubsection{\alg\ vs. Topological} 
We first analyze the performance comparison between \alg\ based method and topological feature based methods
(first five bars).  The best of the topological feature based methods have a
PRAUC value of 0.30, 0.22, 0.137 and 0.14 in Enron, Collaboration, Facebook1,
and Facebook2 dataset (see Figures
\ref{fig:CompareAll}(a), \ref{fig:CompareAll}(c), \ref{fig:CompareAll}(e) and
\ref{fig:CompareAll}(g)), whereas the corresponding PRAUC values for \alg\ are
0.531, 0.362, 0.308, and 0.27, which translates to 77\%, 65\%, 125\%, and
93\% improvement of PRAUC by \alg\ for these datasets.  Superiority of \alg\
over all the topological feature based baseline methods can be attributed to
the capability of Neighborhood based feature representation to capture temporal
characteristics of local neighborhood. Similar trend is observed using 
$NDCG_{50}$ metric, see Figures \ref{fig:CompareAll}(b),
\ref{fig:CompareAll}(d), \ref{fig:CompareAll}(f) and \ref{fig:CompareAll}(h).

\subsubsection{\alg\ vs. Time-Series} 
The performance of time-series based
method (four green bars) is generally better than the topological feature based
methods. The best of the time-series based method has a PRAUC value of 0.503,
0.28, 0.19, and 0.19 on these datasets, and \alg's PRAUC values are better than
these values by 6\%, 29\%, 62\%, and 42\% respectively.  Time-series
based methods, though model the temporal behavior well, probably fail to
capture signals from the neighborhood topology of the node-pairs. Superiority of
\alg\ over Time-Series methods is also similarly indicated by information retrieval
metric $NDCG_{50}$.

\subsubsection{\alg\ vs. DeepWalk} 
The DeepWalk based method (black bars in Figure \ref{fig:CompareAll}) performs much poorly
in terms of both PRAUC and $NDCG_{50}$---even poorer than the topological based method in all four
datasets. Possible reason could be the following: the latent encoding of nodes
by DeepWalk is good for node classification, but the cross-product of those
codes fails to encode the information needed for effective link prediction.

\subsubsection{\alg\ vs. CANDECOMP/PARAFAC (CP)} 

Finally, the tensor factorization based method CP performs marginally better
(around 5\% in PRAUC, and 6\% in $NDCG_{50}$) than \alg\ in small and simple
networks, such as Enron (see Figure \ref{fig:CompareAll}(a, b)). 
But its
performance degrades on comparatively large and complex networks, such as Collaboration, Facebook1
and Facebook2. On Facebook networks, the performance of CP is
even worse than the time-series based methods (see Figures \ref{fig:CompareAll}(e)
and \ref{fig:CompareAll}(g)). \alg\ comfortably outperforms CP on larger
graphs, see Figures \ref{fig:CompareAll}(c, d, e, f, g, h). In terms of PRAUC,
\alg\ outperforms CP by 28\%, 94\%, and 120\% for Collaborative, Facebook1 and Facebook2 networks respectively.
This demonstrates the superiority of \alg\ over one of the best state-of-the-art
dynamic link prediction. A reason for CP's bad
performance on large graphs can be its inability to capture network structure 
and dynamics using high-dimensional tensors representation. 

\subsubsection{Performance across datasets}
When we compare the performance of all the methods across different
datasets, we observe varying performance. For example, for both the metrics,
the performance of dynamic link prediction on Facebook graphs are lower than the
performance on Collaboration graph, which, subsequently, is lower than the
performance on Enron graph, indicating that link prediction in Facebook data is
a harder problem to solve. In these harder networks, \alg\ perform substantially
better than all the other competing methods that we consider in this experiment.

\begin{figure}[H]
\centering
\subfloat[Collaboration Network]{%
\includegraphics[width=.5\linewidth]{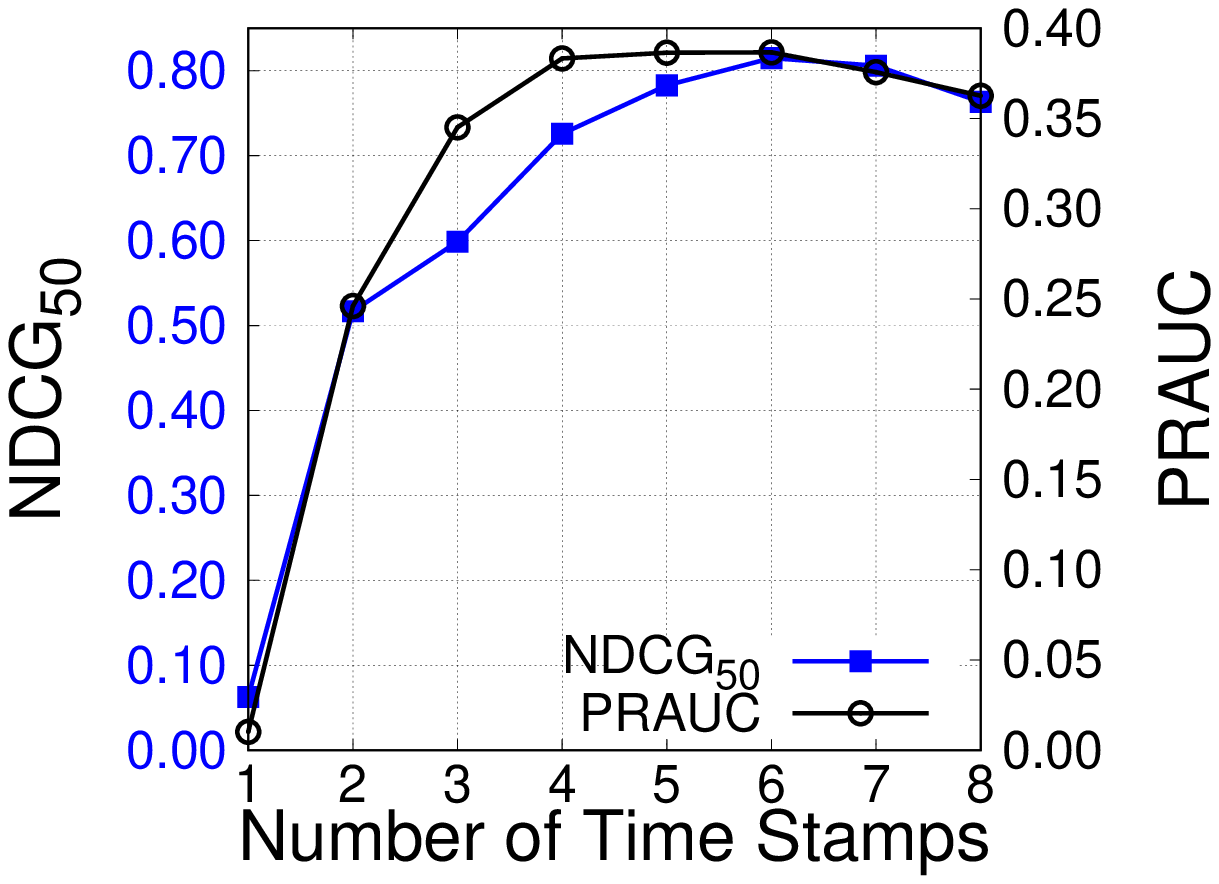}
}
\subfloat[Facebook1 Network]{%
\includegraphics[width=.5\linewidth]{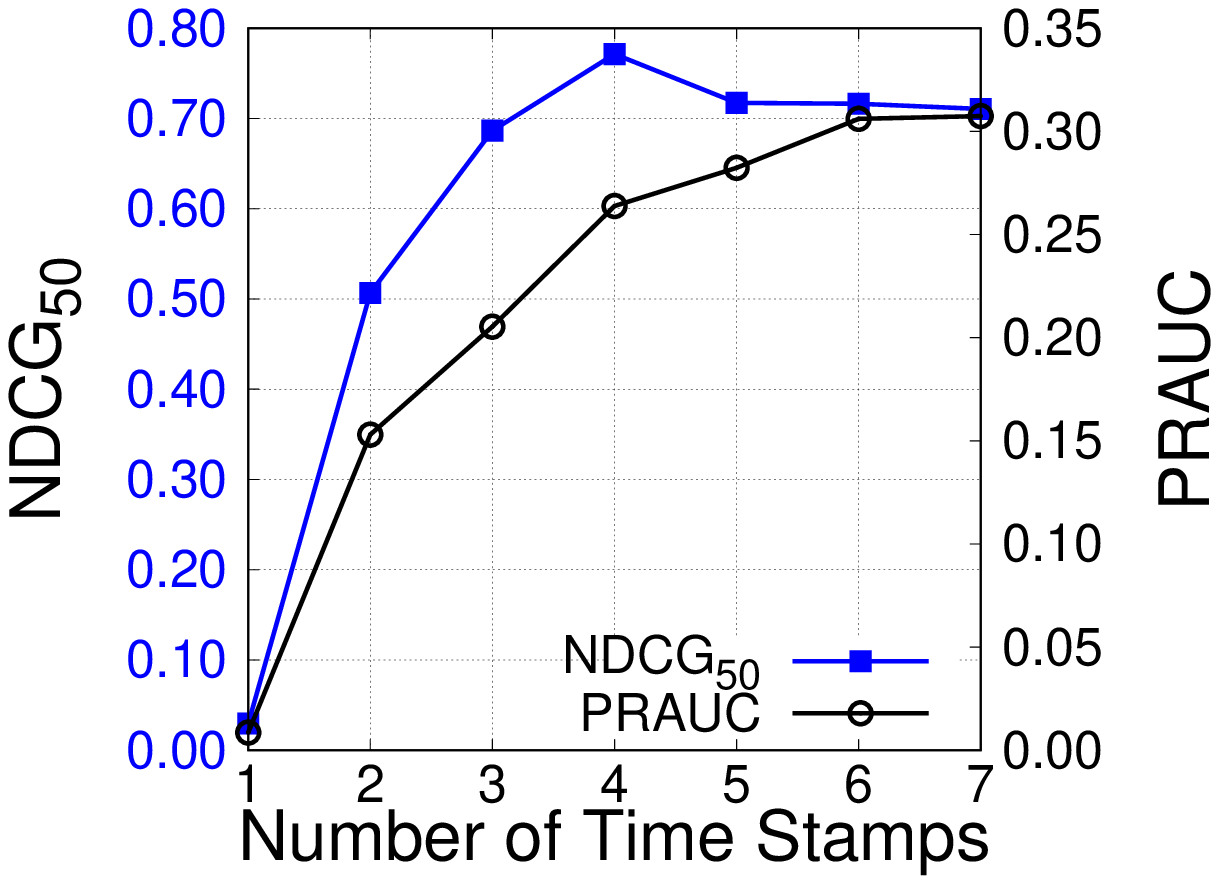}
}
\caption{Change in link prediction performance with number of time stamps. X-axis represents size of training window used
for link prediction.  Largest possible window size depends on number of time
stamps available for the dataset.
}
\label{fig:TimeWindow}
\vspace{-0.3in}
\end{figure}

\subsection{Performance with varying length of Time Stamps}

Besides comparing with competing methods, we also demonstrate the performance
of \alg\ with varying number of available time snapshots.  For this
purpose, we use \alg\ with different counts of past snapshots. For example,
Collaboration dataset has 10 time stamps. The task is to predict links at time
stamp 10.  The largest number of past snapshots we can consider for this data
is 8, where $\mathbf{\widehat{E}}$ is constructed using time stamps $[1-8]$,
and $\mathbf{\overline{E}}$ is constructed using time stamps $[2-9]$. The
smallest number of time stamps we can consider is 1, where
$\mathbf{\widehat{E}}$ is constructed using $[8-8]$, and
$\mathbf{\overline{E}}$ is constructed using $[9-9]$.
In this way, by varying the length of historical time stamps, 
we can evaluate the effect of time stamp's length on the performance of a link prediction method.

The result is illustrated in Figure \ref{fig:TimeWindow}. The x-axis represents
the number of time stamps used by \alg , the left y-axis represents the
$NDGC_{50}$ and the right y-axis represents the
PRAUC.  Figures \ref{fig:TimeWindow}(a),
and \ref{fig:TimeWindow}(b) corresponds to the results obtained on Collaboration and Facebook1, respectively.

We observe from Figure \ref{fig:TimeWindow} that the performance ($NDGC_{50}$
and PRAUC) of link prediction increases with increased number of time stamps.
But beyond a given number of snapshots, the performance increment becomes
insignificant. The performance starts to deteriorate after certain number of snapshots (see Figure \ref{fig:TimeWindow}(a)).  This may be because of the added complexity of the optimization framework with increased number of time stamps.  We also observe
consistent improvement of performance with the number of snapshots for the
Facebook1 data (Figure \ref{fig:TimeWindow}(b)),
which indicates that for this dataset link information from distant
history is useful for dynamic link prediction. We do not show results of Enron and Facebook2  for this experiment, 
because of space constraint, however, they show similar trends.

\begin{figure}[H]
\centering
\includegraphics[width=.8\linewidth]{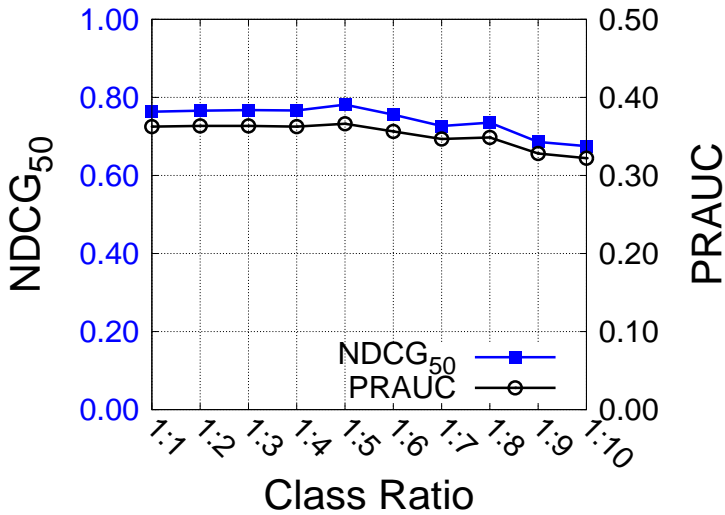}
\caption{Effect of class imbalance in link prediction performance on Collaboration network.}
\label{fig:classImbalance}
\vspace{-0.3in}
\end{figure}

\subsection{Effect of Class Imbalance on Performance}
In link prediction problem, class imbalance is a prevalent issue. The class imbalance problem appears in a classification task, when the dataset contains imbalanced number of samples for different classes. In link prediction problem, the number of positive node-pairs (with an edge) is very small compared to the number of negative node-pairs (with no edge), causing class imbalance problem. 

To demonstrate the effect of class imbalance in link prediction task, we perform link prediction using \alg\ embeddings with different level of class imbalance in the training dataset.
We construct the training dataset by taking all positive node-pairs and sampling from the set of negative node-pairs. 
For a balanced dataset, the number of negative samples will be equal to the number of all positive node-pairs considered. Thus, the balanced training dataset has positive node-pairs to negative node-pairs ratio $1:1$. At this point, the only way to increase the size of the data is to increase the sample size for negative node-pairs. Consequently, the ratio of classes also increases towards negative node-pair.
Figure \ref{fig:classImbalance} shows gradual decrease in link prediction performance in Collaboration network with the increase of imbalance (see ratios in X-axis) in the dataset (despite the fact that the dataset gets larger by adding negative node-pairs). 

This result advocates towards the design choice of under-sampling \cite{Lichtenwalter:2010} of negative node-pairs by uniformly sampling from all negative node-pairs, so that the training set has equal numbers of positive and negative node-pairs. Under-sampling, helps
to mitigate the problem of class imbalance while also reducing the size of the training dataset.

\section{Conclusion}\label{sec:conclusion}
In this paper, we present \alg\ a learning method for obtaining feature representation of node-pairs in dynamic networks.
We also give classification based link prediction method, which uses \alg\ feature representation for future link prediction in dynamic network setup.
The proposed link prediction method
outperforms several existing methods that are based on topological
features, time series, deep learning and tensor analysis.


\bibliographystyle{spmpsci}
\bibliography{dylink2vec}


\end{document}